\documentclass[%
 reprint,
 amsmath,amssymb,
 aps,
 pra,
floatfix,
]{revtex4-2}

\usepackage{graphicx}
\usepackage{dcolumn}
\usepackage{bm}
\usepackage{braket}
\usepackage{xcolor}

\begin{document}

\preprint{APS/123-QED}

\title{Sub-radiant states for imperfect quantum emitters coupled by a nanophotonic waveguide}

\author{Xiao-Liu Chu}
 \altaffiliation[Present address: ]{MRC London Institute of Medical Sciences, Du Cane road, London, W12 0NN, United Kingdom}
 \email{xchu@ic.ac.uk}
 \author{Vasiliki Angelopoulou}
 \affiliation{%
 Center for Hybrid Quantum Networks (Hy-Q), Niels Bohr Institute,
University of Copenhagen, Blegdamsvej 17, DK-2100 Copenhagen, Denmark\\
}%
 \author{Peter Lodahl}
 \author{Nir Rotenberg}%
 \altaffiliation[Present address: ]{Centre for Nanophotonics, Department of Physics, Engineering Physics \& Astronomy, 64 Bader Lane, Queen’s University, Kingston, Ontario, Canada K7L 3N6}
\affiliation{%
 Center for Hybrid Quantum Networks (Hy-Q), Niels Bohr Institute,
University of Copenhagen, Blegdamsvej 17, DK-2100 Copenhagen, Denmark\\
}%


\date{\today}

\begin{abstract}
Coherent interactions between quantum emitters in tailored photonic structures is a fundamental building block for future quantum technologies, but remains challenging to observe in complex solid-state environments, where the role of decoherence must be considered. Here, we investigate the optical interaction between two quantum emitters mediated by one-dimensional waveguides in a realistic solid-state environment, focusing on the creation, population and detection of a sub-radiant state, in the presence of dephasing. We show that as dephasing increases, the signatures of sub-radiance quickly vanish in intensity measurements yet remain pronounced in photon correlation measurements, particularly when the two emitters are pumped separately so as to populate the sub-radiant state efficiently. The applied Green's tensor approach is used to model a photonic crystal waveguide, including the dependence on the spatial position of the integrated emitter. The work lays out a route to the experimental realization of sub-radiant states in nanophotonic waveguides containing solid-state emitters.

\end{abstract}

\maketitle


\section{\label{sec:introduction}Introduction}

With the maturation of quantum photonic platforms, light-matter interaction is no longer limited to the control of single emitters \cite{Lodahl:15, Schmidgall:18,Turschmann:19}, as multi-emitter systems are increasingly being developed and investigated \cite{Sipahigil:16,Kim:18,Grim:19,Trebbia:22}. Of the different platforms, waveguide quantum electrodynamics (w-QED) where quantum emitters are coupled to photonic waveguides \cite{Lundhansen:08,Sipahigil:16,Turschmann:17b,Kim:18,Grim:19,Pennetta:22}, is particularly promising. Solid-state w-QED systems allow on-chip integration based on reliable nanofabrication \cite{Burek:17,Turschmann:17b,Pregnolato:20}, permitting electrical control over the emitter resonances \cite{Turschmann:17,Schmidgall:18,Thyrrestrup:18}, although scaling-up to many emitters is challenged by inhomogeneous broadening unlike the case of atoms \cite{Pennetta:22}. Importantly, even few quantum emitters deterministically coupled to a waveguide can be a very powerful quantum resource since each emitter can produce a high number of photonic qubits. Indeed two coherently-coupled quantum emitters have been proposed as sufficient for creating large-scale photonic cluster states for quantum communication \cite{segovia:19,Kimble:08} or to access decoherence-free sub-spaces for quantum computation \cite{Beige:00}. The latter requires the creation of long-lived sub-radiant states: collective excitations with lifetimes that are much longer than those of the single emitters.

Previous studies have focused on systems of ideal emitters \cite{Dzsotjan:10,Asenjogarcia:17,Das:18}, while here we show how sub-radiant states may be created and populated using real-world quantum emitters in the presence of imperfections. We specifically consider solid-state emitters \cite{aharonovich:16} coupled to nanophotonic waveguides and focus on the role of pure dephasing. We show that while even a small amount dephasing (less than $10\%$ of the emitter decay rate $\Gamma_0$) is sufficient to remove the signature of sub-radiance in photonic intensity measurements, the states can be efficiently excited and detected from photon correlation measurements. 
We observe that dephasing introduces a mixing between the super and sub-radiant states, providing another pathway to populate states that would otherwise be inaccessible.  Finally, we calculate the Green's tensor for a photonic crystal waveguide and apply that to determine the spatial dependence of the emitter-emitter coupling.
\begin{figure}
\includegraphics[trim={0 0cm 0 0},clip, width=0.45\textwidth]{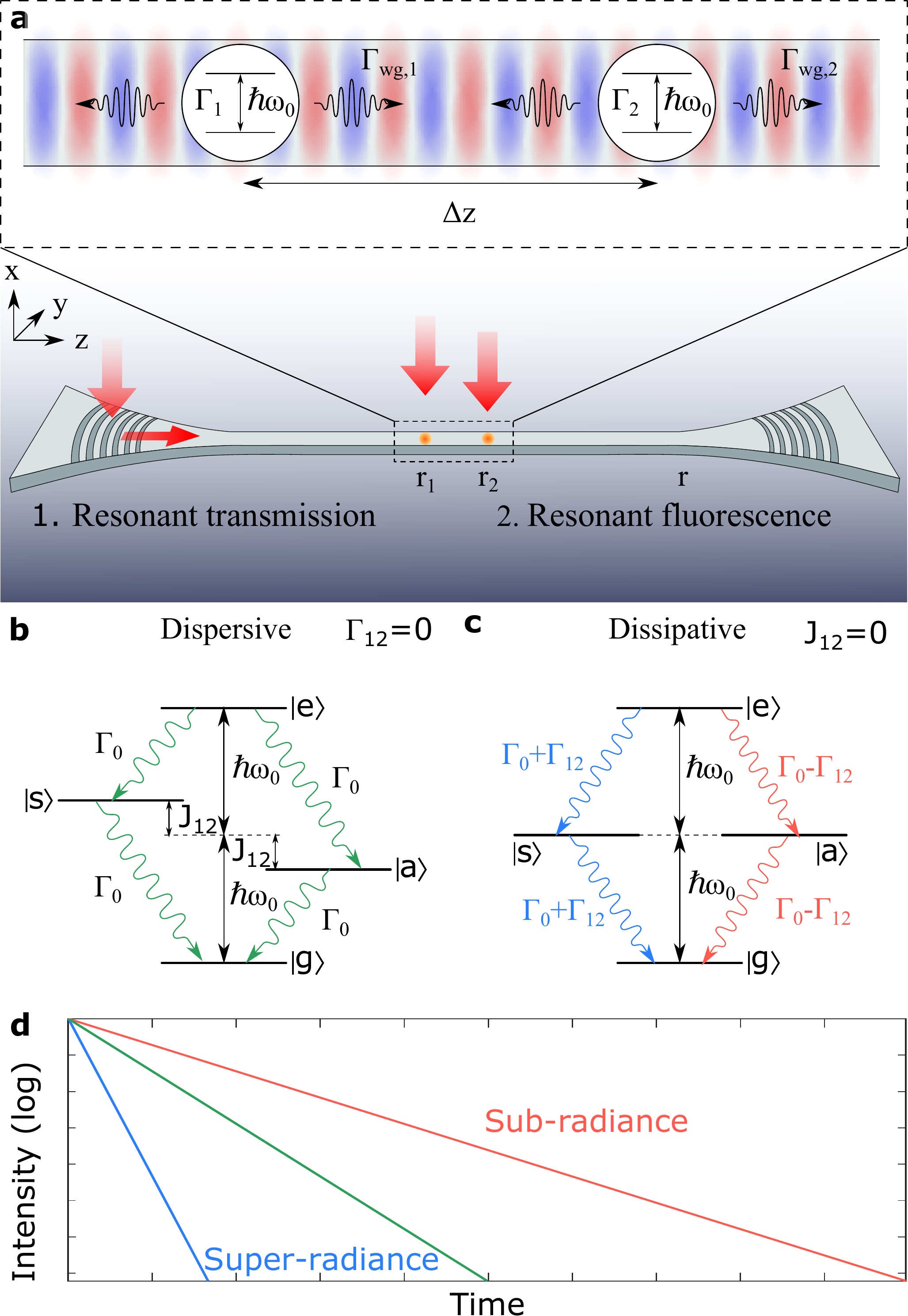}
\caption{\label{fig:energylevel}a. Two spatially separated emitters (by $\Delta z$) couple to the same electromagnetic field of a single waveguide mode (See inset). This coupled system may be excited either through the waveguide mode (resonant transmission) or from free-space (resonant fluorescence), in which case the emitters may be addressed individually. The resultant four energy levels of the coupled system when the coupling is b) fully dispersive or c) fully dissipative, along with d) exemplary time-dependent emission traces from the different transitions. For dispersive coupling, the decay rates of the coupled transitions remain the same as that of the individual systems, corresponding to the green curve in d), while the energy levels are shifted. In contrast, for dissipative coupling the energy levels are unchanged while the decay rates are altered by $\Gamma_{12}$ such that one transition is sub-radiant (shown in red here and in d) and on is super-radiant (shown in blue here and in d). Depending on emitter position, the coupling may be partially dispersive and partially dissipative.}
\end{figure}

\section{\label{sec:geometry} sub-radiance of two quantum emitters in a nanophotonic waveguide}
\subsection{\label{sec:theory}System geometry, excitation and measurements}
The overall goal is to exploit how to efficiently control and populate a sub-radiant state using coupled quantum emitters in a nanophotonic waveguide and study the robustness towards imperfections. The system is sketched in Fig.~\ref{fig:energylevel} and is comprised of two, two-level emitters (TLEs) with identical resonant frequencies $\omega_0 \equiv \omega_1 = \omega_2$, and decay rates $\Gamma_{0} \equiv \Gamma_{1} = \Gamma_{2}$. Both emitters couple to the same guided mode, and are separated by a distance $\Delta z_{12} \gg \lambda_\mathrm{wg}$. Here, $\lambda_\mathrm{wg}$ is the effective wavelength of the guided mode, at $\omega_0$, meaning that the TLEs are coupled via the long-range dipole-dipole interaction mediated by the guided mode \cite{Asenjogarcia:17}. 

When coupled, excitations are no longer ascribed to individual emitters, but rather the system as a whole, resulting in the Dicke energy states as shown in Fig.~\ref{fig:energylevel}b and c. These coupled states are \cite{Dicke:54}:
\begin{eqnarray}
&\ket{g}=\ket{g_1}\otimes\ket{g_2},\nonumber\\
&\ket{s} =\frac{1}{\sqrt{2}}(\ket{e_1}\otimes\ket{g_2}+\ket{g_1}\otimes\ket{e_2}),\nonumber \\
&\ket{a} =\frac{1}{\sqrt{2}}(\ket{e_1}\otimes\ket{g_2}-\ket{g_1}\otimes\ket{e_2}), \nonumber \\
&\ket{e} =\ket{e_1}\otimes\ket{e_2}.\label{eq:Dicke}
\end{eqnarray}
Not only are the intermediate states $\ket{s}$ and $\ket{a}$ entangled \cite{Almutairi:11,Martincano:11}, but both the decay rates and energy levels are modified by $\Gamma_{12} = \Gamma_{21}$ and $J_{12} = J_{21}$, respectively, as shown in the energy diagram (the full expression for these terms is detailed in the next section). In what follows, we focus on identifying and populating the sub-radiant state, whose decay rate $\Gamma_{\mathrm{s/a}}=\Gamma_{0} \pm \Gamma_{12} \ll \Gamma_{0}$. In the case where $\Gamma_{12}$ is a positive quantity (negative), the asymmetric state $\ket{a}$ is sub-radiant (and $\ket{s}$ super-radiant), and vice versa for a negative $\Gamma_{12}$.

There are two ways to excite the w-QED system, both of which are shown in Fig.~\ref{fig:energylevel}a. Either both emitters are excited through the waveguide in the resonant transmission (RT) configuration, or they can be addressed through free-space in the resonant fluorescence (RF) configuration. For RT, the relative phases with which the emitters are excited is locked by the properties of the guided mode to which they couple, and determined by their separation. In this configuration, the transmitted signal includes pump photons that coherently interfere with those scattered from the emitters \cite{Turschmann:19}. In contrast, free-space excitation allows either, or both, of the emitters to be addressed with any arbitrary phase or intensity difference. Here, only photons that originate from the emitters are detected.

Regardless of the excitation method, we detect the light field at the output port of the waveguide, as shown in Fig.~\ref{fig:energylevel}a. In practice this requires out-coupling of the light, for example using a grating as shown in the figure, while we calculate the field in the waveguide far away from the emitters. We calculate the intensity spectrum in either of the two configurations,
\begin{subequations}
\begin{eqnarray}
\text{I}_{\mathrm{RT}}\left(\bf r, \omega\right) &= \frac{\braket{\hat{\bf{E}}^-\left(\bf r, \omega\right)\hat{\bf{E}}^+\left(\bf{r}, \omega\right)}}{\braket{\hat{\bf{E}}_{\text{p}}^-\left(\bf{r}, \omega\right)\hat{\bf{E}}_{\text{p}}^+\left(\bf{r}, \omega\right)}},  \label{eq:I_RT}\\
\text{I}_{\mathrm{RF}}\left(\bf{r}, \omega\right) &= \frac{\braket{\hat{\bf{E}}^-_{\text{scat}}\left(\bf{r}, \omega\right)\hat{\bf{E}}^+_{\text{scat}}\left(\bf{r}, \omega\right)}}{\braket{\hat{\bf{E}}_{\text{p}}^-\left(\bf{r}, \omega\right)\hat{\bf{E}}_{\text{p}}^+\left(\bf{r}, \omega\right)}}. \label{eq:I_RF}
\end{eqnarray}
\end{subequations}
or the time-resolved second-order correlation function,
\begin{equation}
    g^{(2)}(t,\tau)=\frac{\langle \hat{\bf{E}}^-(t) \hat{\bf{E}}^-(t+\tau) \hat{\bf{E}}^+(t+\tau) \hat{\bf{E}}^+(t) \rangle}{\langle \hat{\bf{E}}^-(t) \hat{\bf{E}}^+(t)  \rangle^2}. \label{eq:g2}
\end{equation}
In these equations, the electric-field operator can be expressed either in frequency or time, using the standard Fourier transform relations. In either case, the total field operator is comprised of a positive and negative frequency component $\hat{\bf{E}}=\hat{\bf{E}}^+ + \hat{\bf{E}}^-$. Furthermore, each component of the total field can be written as a sum of the excitation field $\hat{\bf{E}}_{\text{p}}^{\pm}$ and the field scattered (in the low-excitation regime, as is the case in this work) or emitted (for strong excitations) from the emitter $\hat{\bf{E}}^{\pm}_{\text{scat}}$. In the case of RF excitation, only the scattered field arrives at the detection point.

\subsection{\label{sec:wgtheory}Waveguide-coupled emitters}
Of the different approaches that model a system of coupled quantum emitters \cite{Kien:05,Zheng:13,Asenjogarcia:17,Das:18, Regidor:21}, we employ a Green's tensor formalism developed to describe dipolar emitters in realistic nanophotonic systems that are dispersive, dissipative and accounting for pure dephasing of the embedded emitters \cite{Dung:02}. Previously, this formalism has been employed to model the coupling of two quantum emitters via a plasmonic channel \cite{Dzsotjan:10,Martincano:10,Martincano:11,Gonzaleztudela:11}, and extended to systems of $N$ emitters coupled through a nanophotonic channel \cite{Fang:14,Haakh:16,Asenjogarcia:17,Das:18}, in all cases assuming perfect coherence. We extend this theory to include the effect of pure-dephasing for coupled emitters, summarizing the basic principles of the theory and the way in which we account for this decoherence.

The single emitter dipole projected Green's function, $ \mathrm{G}(\textbf{r},\textbf{r}_1) = \frac{i}{2}\Gamma_1\beta_1 e^{ik(z-z_1)}$ \cite{Asenjogarcia:17}, can be generalized to two identical, coupled emitters in a one-dimensional nanobeam waveguide such as is sketched in Fig.~\ref{fig:energylevel}a and can be written as
\begin{eqnarray}
\mathrm{G}(\bf{r}_1,\bf{r}_2,\omega_{\text{p}}) &= i \Gamma_0 \sqrt{\beta_1\beta_2} \frac{\hbar}{2\mu_0\omega_p^2}e^{ik|\Delta z_{12}|},
\label{eq:Green1D}
\end{eqnarray}
where $\omega_p$ is the optical angular frequency, $\mu_0$ is the permeability of free-space and $k=2 \pi / \lambda_{\mathrm{wg}}$ is the wavenumber of the optical mode, whose effective wavelength is $\lambda_{\mathrm{wg}}$. Due to the quasi-1D nature of the waveguide, the phase of the Green's function only depends on the emitter separation $\Delta z$, while the emission rate $\Gamma_0$ and emitter-waveguide coupling efficiencies $\beta_i = \Gamma_{\mathrm{wg,}i} / \Gamma_0$ depend implicitly on the positions of the emitters $\bf{r}_1$ and $\bf{r}_2$.

Knowledge of this Green's function allows us to calculate both the electric field operator and the dispersive and dissipative coupling terms. If we assume transverse ($\hat{\bf{y}}$-oriented) transition dipole moments $\bf{d}_1$ and $\bf{d}_2$, then the positive electric field operator isgiven by~\cite{Chen:13,Hood:17},
\begin{eqnarray}
\hat{\bf{E}}^+_{\text{y}}(\bf{r}) 
&=e^{ikz} (\hat{\bf{E}}^+_{\text{p0,y}}+\frac{i\hbar\Gamma_1\beta_1e^{-ikz_1} }{2} \hat{\sigma}_{ge}^1 \nonumber \\
&+\frac{i\hbar\Gamma_2\beta_2e^{-ikz_2} }{2} \hat{\sigma}_{ge}^2 ),
\label{eq:Ey}
\end{eqnarray}
with the negative component given by the Hermitian conjugate of this expression. Here, $\hat{\sigma}_{ge}^{i} = \ket{g_i}\bra{e_i}$ is the lowering operator for emitter $i$. As is outlined in Appendix~\ref{appen:sec1}, these emitter operators are analysed in the full system Hamiltonian and contain information about the emitter coupling rate, as well as the effects of dephasing and excitation strength $\Omega_{p,i}$. We identify the first term in Eq.~\ref{eq:Ey} as the incident pump field (which is only present in the RT configuration), while the following two terms represent the fields scattered from emitters 1 and 2, respectively. In RF and resonant reflection (RR), the same scattered field terms are used, except in RF the relative phase of the pump fields can be tuned independently.

Within this formalism, the dispersive and dissipative coupling terms as seen in Fig.~\ref{fig:energylevel} are,~\cite{Asenjogarcia:17,Hood:17,Turschmann:17,Gonzaleztudela:11}
\begin{eqnarray}
J_{12} &= \frac{\mu_0\omega_p^2}{\hbar}\bf{d}^*_1\cdot \text{Re}[\bf{G}(\bf{r}_1,\bf{r}_2,\omega_p)]\cdot\bf{d}_2 \nonumber \\
&=\frac{1}{2}\Gamma_0\sqrt{\beta_1\beta_2}\text{sin}(k\Delta z_{12}),
\label{eq:J12}
\end{eqnarray}	
and,
\begin{eqnarray}
\Gamma_{12} &= \frac{2\mu_0\omega_p^2}{\hbar}\bf{d}^*_1\cdot \text{Im}[\bf{G}(\bf{r}_1,\bf{r}_2,\omega_p)]\cdot\bf{d}_2  \nonumber \\
&= \Gamma_0\sqrt{\beta_1\beta_2}\text{cos}(k\Delta z_{12}),
\label{eq:Gamma12}
\end{eqnarray}
respectively, where we have used $\mathrm{G}(\bf{r}_1,\bf{r}_2,\omega_{\text{p}}) = \bf{d}^*_1\cdot \bf{G}(\bf{r}_1,\bf{r}_2,\omega_p)\cdot\bf{d}_2$. As the distance between the emitters varies, the coupling changes between being dispersive and dissipative, where the latter results in the long-lived sub-radiant states. Note that in more complex nanophotonic systems, such as the photonic crystal waveguides studied in Sec.~\ref{sec:phcwg}, the Green's tensor is much more complex and Eqs.~(\ref{eq:J12}) and~(\ref{eq:Gamma12}) are modified.
\begin{figure*}
    \includegraphics[trim={0 0cm 0 0},clip, width=0.9\textwidth]{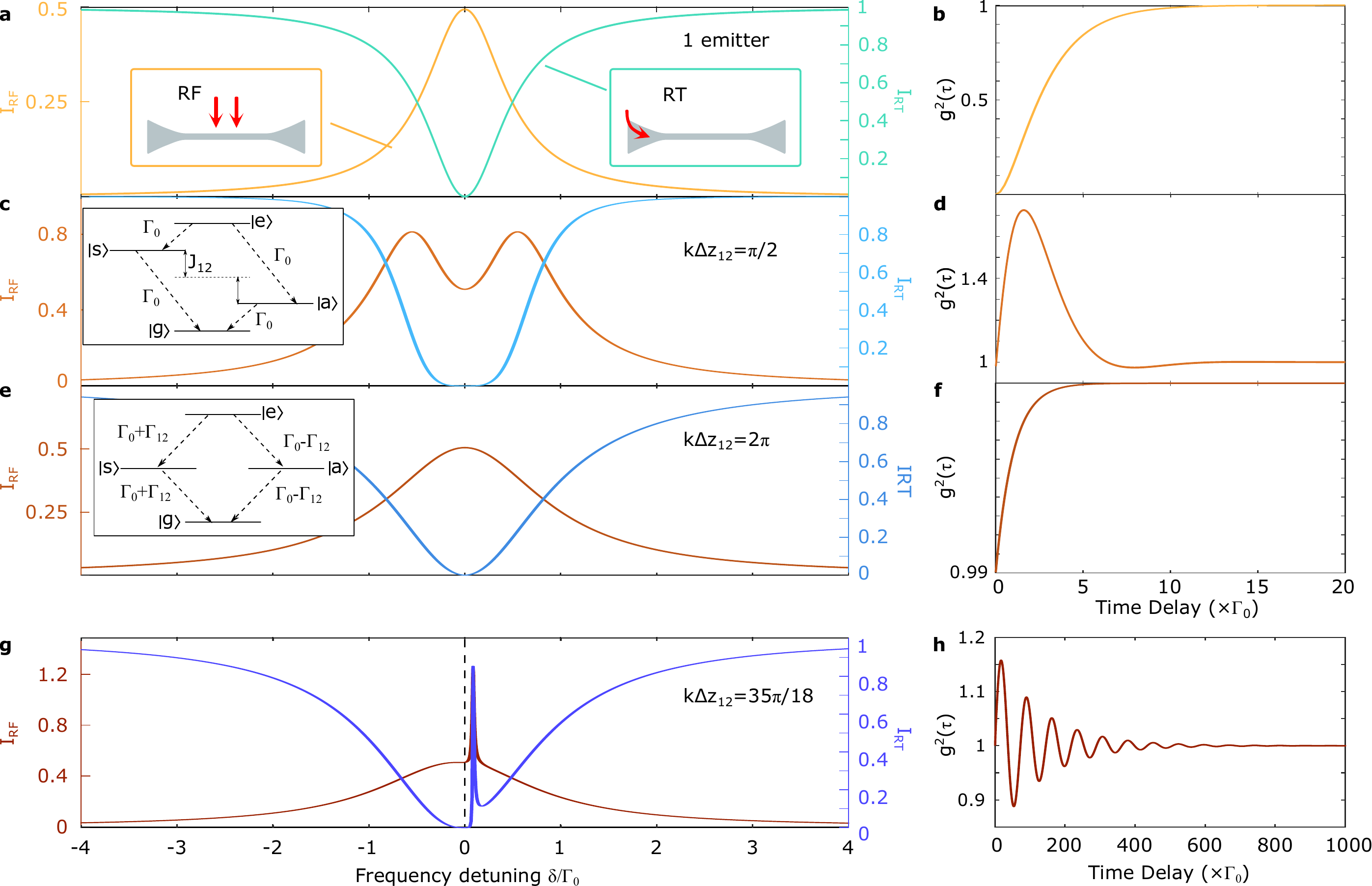}
    \caption{\label{fig:ideal} Ideal system response in the low-power limit $\Omega_{p,i}=10^{-7}$. The RF (left axis) and RT (right axis) for a) a single emitter in a waveguide, and two emitters separated by an effective phase of c) $\pi/2$, e) $2\pi$ and g) $35\pi/18$. $g^{(2)}(\tau)$ of the corresponding RF  on resonance $\left(\delta = 0 \right)$ are shown in b) - h). c) and d) correspond to ideal dispersive coupling where the energy levels are shifted but the decay rate is unaltered, while e) and f) represent ideal dissipative coupling where only the decay rate is altered. In the latter case, only the super-radiant state is populated, as reflected by the rapid rise of the correlation function. Slightly changing the relative-phase, as shown in g) and h), enables the population of the sub-radiant state, whose decay rate is now finite.}
\end{figure*}

Using Eqs.~(\ref{eq:J12}) and~(\ref{eq:Gamma12}), we solve the Master's Equation of the full system Hamiltonian and Lindblad superoperator which includes the pure-dephasing term, as detailed in Appendix~\ref{appen:sec1}. We recover the atomic operators $\hat{\sigma}_{ge}^1$ and $\hat{\sigma}_{ge}^2$ as required for Eq.~(\ref{eq:Ey}) to determine the population of the different states. We extract both the steady-state values, as required for the intensity calculations, and time-dependent expectation values of atomic operator products, which is required for the photon-correlation function $g^{(2)}(t,\tau)$. The time dependence is obtained by the use of the quantum regression theorem \cite{Lax:63,tannoudji:04,Hobson:11}.

As a baseline, we first consider the ideal case where there is no dephasing (i.e. $\Gamma_{\mathrm{dp}} = 0$), unity radiative coupling $\beta_i=1$, $i=1,2$ and at low pumping strength $\Omega_{p,i}=10^{-7}$. The results in Fig.~\ref{fig:ideal} are compared to results from the literature \cite{Fang:14,Das:18}. The case of RF of a single emitter driven by a continuous-wave field is shown in Fig.~\ref{fig:ideal}a (left axis) and the resonance peak corresponds to single photons being emitted into the waveguide. Furthermore, the RT spectral response (right axis) displays a dip due to the destructive interference between the forwardly emitted photons and the incoming pump field~\cite{Turschmann:19}. The photon correlation function  $(g^{(2)}(\tau))$ displays the well-known anti-bunching behavior of a single-photon emitter, cf. Fig.~\ref{fig:ideal}b, with an internal recovery-time of the emitter lifetime ($\Gamma_0^{-1}$).

When the effective phase is $k\Delta z_{12}  = (m+1/2) \pi $ or $k\Delta z_{12}  = p \pi$ for $m, p$ integers,  the coupling is purely dispersive or dissipative, as shown in Figs.~\ref{fig:ideal}c and e, respectively. For an ideal dispersive coupling, $J_{12} = \Gamma_0$ and $\Gamma_{12} = 0$, meaning that the energy degeneracy of the two transitions is lifted. In RT, this results in the appearance of a broad resonance, where the individual dips are not individually resolved. In contrast, the RF spectrum displays two clear peaks, and  the corresponding $g^{(2)}(\tau)$ measured between them at $\delta = 0$ (Fig.~\ref{fig:ideal}d) is oscillatory due to the quantum interference between the two transitions. For RF, the emitters are excited with a relative phase difference corresponding to the separation, $k\Delta z_{12}$.

For ideal dissipative coupling (Fig.~\ref{fig:ideal}e and f) $J_{12} = 0$ and $\Gamma_{12} = \Gamma_0$, meaning that both the sub-radiant $\ket{a}$ and super-radiant $\ket{s}$ transitions are available. However, both RT and RF spectra are determined purely by the super-radiant transitions that are broadened compared to the single emitter spectra of a). This is confirmed by the $g^{(2)}(\tau)$ trace, which evolves more rapidly. The perfect sub-radiant state $(\Gamma_0 - \Gamma_{12} = 0)$ cannot emit, and therefore does not contribute to the output spectra.

Sub-radiant behaviour may be observed when deviating from the situation of ideal dissipative coupling ($k\Delta z_{12}  = 2 \pi$). Figures~\ref{fig:ideal}g and h show $k\Delta z_{12}  = 35 \pi / 18$, such that the sub-radiant state has a finite lifetime. In this case, we observe a sharp, slightly detuned feature corresponding to the sub-radiant state in both RF and RT co-existing with the broadened super-radiant resonance. This interpretation is confirmed by the $g^{(2)}(\tau)$ trace, where oscillations due to interference between the two states decay on time scales of 100's of $\Gamma_0^{-1}$. Further details on the dependence of the sub-radiant peak position and width, as well as the decay rate as seen in the $g^{(2)}(\tau)$ calculations, on the relative phase between the emitters are given in Appendix~\ref{appen:sec2}.

\subsection{\label{sec:imperfect_emit} Role of dephasing and imperfect radiative coupling}
\begin{figure}
   \includegraphics[trim={0 0cm 0 0},clip, width=0.48\textwidth]{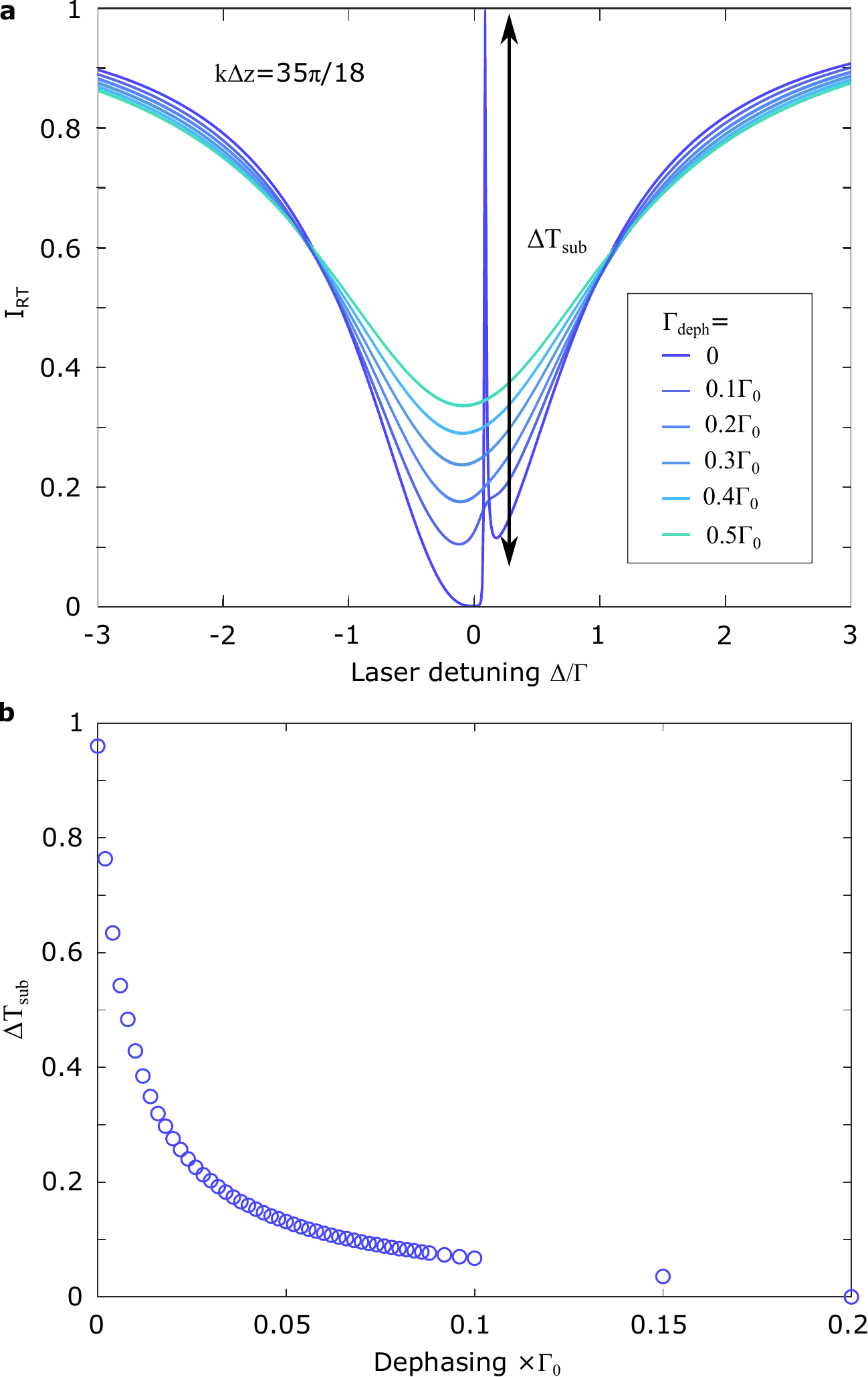}
   \caption{\label{fig:RT_deph} sub-radiant features in intensity RT in the presence of dephasing. a) $I_\mathrm{RT}$ spectra for different $\Gamma_{\mathrm{deph}}$ for $k\Delta z_{12} = 35 \pi / 18$. The sharp sub-radiant feature, of magnitude $\Delta T_\mathrm{sub}$, rapidly vanishes as $\Gamma_{\mathrm{deph}}$ increases, as shown in b). }
\end{figure}

We now consider the case where dephasing is present. Fig.~\ref{fig:RT_deph} shows frequency-dependent RT  when $\Gamma_\mathrm{deph}$ ranges between 0 and $0.5\Gamma_0$ (see Appendix~\ref{appen:sec2} for the corresponding reflections).
\begin{figure}
    \includegraphics[trim={0 0cm 0 0},clip, width=0.45\textwidth]{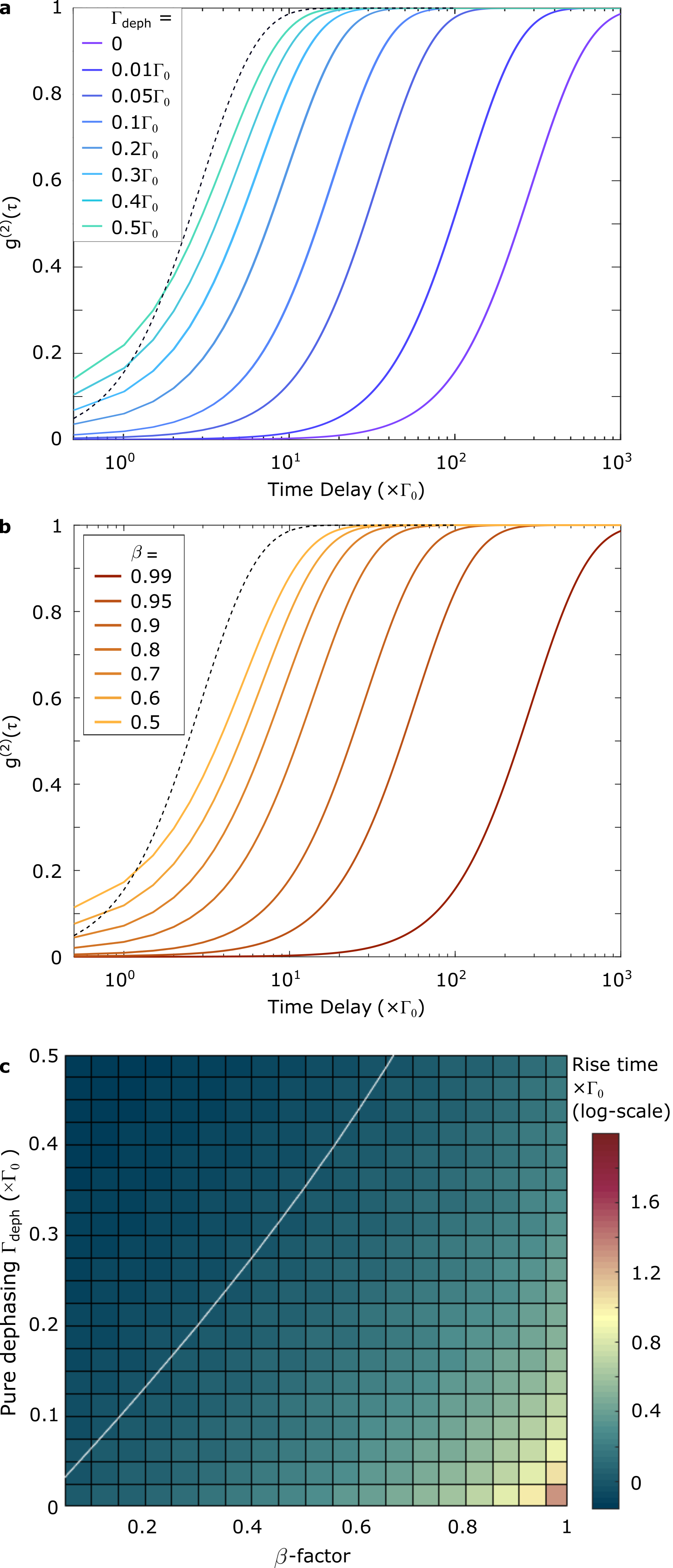}
  \caption{\label{fig:RF_g2} The observed $g^{(2)}(\tau)$ of the coupled emitter system when the emitter separation is $\Delta z_{12} = \pi/k$ and the emitters are probed in RF, where the pump fields are in phase, resulting in the excitation of the symmetric (sub-radiant) state. The observed antibunching has a slow rise time that varies with a) dephasing $\Gamma_{deph}$ and b) $\beta$ ($=\beta_1=\beta_2$). For comparison, the ideal, single-emitter $g^{(2)}(\tau)$ trace (c.f. Fig.~\ref{fig:ideal}b) is shown by the dashed curve. c) 2D map of the rise time as a function of $\Gamma_{deph}$ and $\beta$.}
\end{figure}

The super-radiant component of the RT spectra, shown in Fig.~\ref{fig:RT_deph}a, broadens and becomes shallower as $\Gamma_{\mathrm{deph}}$ increases, as compared to the single emitter case. Similarly, the sharp peak indicating the existence of a sub-radiant state, $\Delta T_\mathrm{sub}$, rapidly vanishes as $\Gamma_{\mathrm{deph}}$ increases, as can be seen in Fig.~\ref{fig:RT_deph}b, where we plot the magnitude of the sub-radiant feature, $\Delta T_\mathrm{sub}$, as a function of $\Gamma_{\mathrm{deph}}$. For state-of-the-art values of $\Gamma_{\mathrm{deph}} = 0.016 \Gamma_0$ \cite{jeannic:21}, we expect $\Delta T_\mathrm{sub} = 0.32$. Note that residual spectral diffusion, due to a slow drift present in many experimental implementations, will reduce the sub-radiant feature further.

In contrast, and as we observed in Fig.~\ref{fig:ideal}, the signature of sub-radiance in the time dynamics, such as the photon-correlation function $g^{(2)}(\tau)$, is pronounced and leads to a slow modulation of the trace. Furthermore, in the RF configuration, the individual transitions can be addressed by tuning the relative phase of the beams that excite each emitter, as discussed above. Figure~\ref{fig:RF_g2} shows the time-dependent $g^{(2)}(\tau)$ as a function of a) $\Gamma_{\mathrm{deph}}$ (for almost ideal coupling, $\beta = 0.99$) and b) $\beta$ (when $\Gamma_{\mathrm{deph}} = 0$), for two emitters separated by $\Delta z_{12} = \pi/k$ and where the emitters are pumped in phase, such that the sub-radiant state $\ket{s}$ is excited. (Note that the x-axis does not extend all the way to zero).
\begin{figure*}
    \includegraphics[trim={0 0cm 0 0},clip, width=0.9\textwidth]{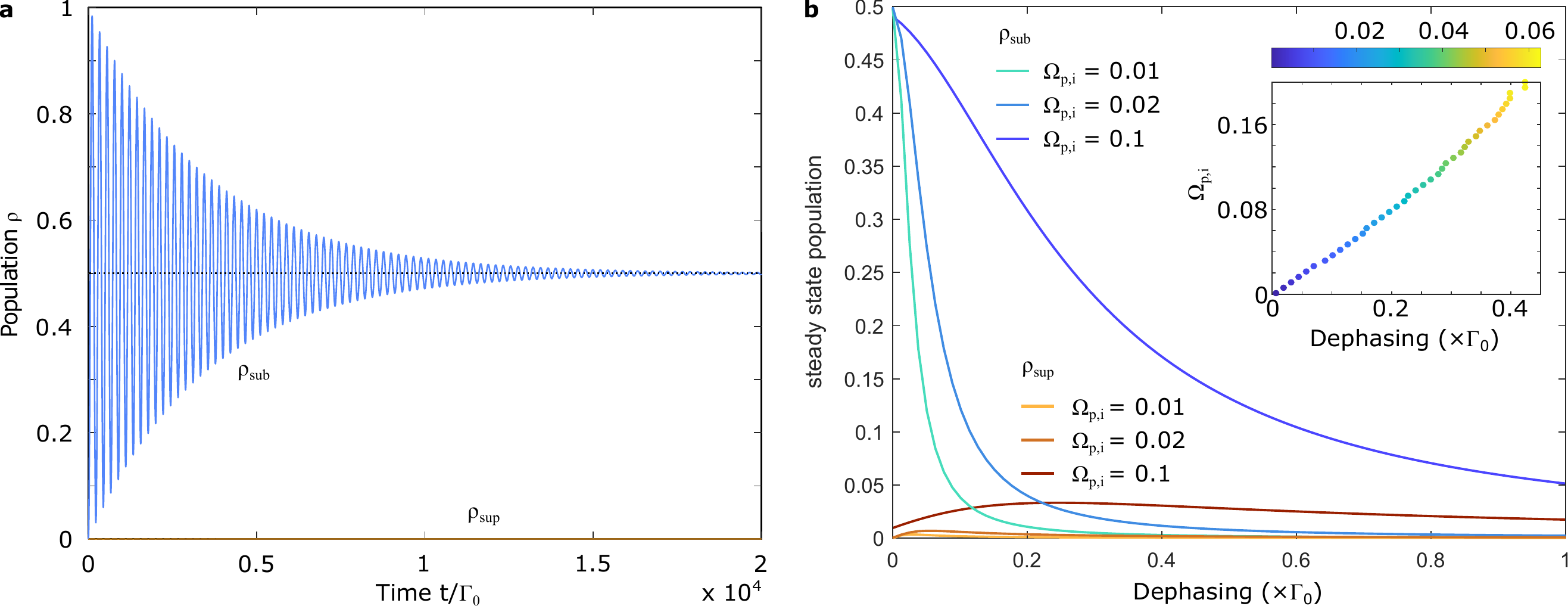}
    \caption{\label{fig:rho} The observed population dynamics of the super- and sub-radiant states, starting with both emitters initially in the ground state, with a separation $\Delta z_{12} = \pi/k$. a) When the emitters are excited in phase ($\Omega_{p,i}= 0.01$), the population of the symmetric (sub-radiant) state reaches a steady state population of 0.5 and the super-radiant state is not populated. b) By introducing pure dephasing, the super-radiant state is also populated as seen in the plotted steady-state values. Inset: maximum steady-state population (given by colorbar) of the super-radiant state for different $\Gamma_{deph}$ and $\Omega_{p,i}$.}
\end{figure*}

As the imperfections of the coupled system increase, the rise-time shortens (see Appendix~\ref{appen:sec3} for definition). For example, when increasing $\Gamma_{\mathrm{deph}}$ from 0 to $0.01\Gamma_0$ to $0.1\Gamma_0$, the rise-time changes from $100\Gamma_0^{-1}$ to $38.9\Gamma_0^{-1}$ to $6.2$ $\Gamma_0^{-1}$ ($\beta$=0.99). Similarly, changing $\beta$ from 0.99 to 0.9 to 0.7 changes the rise-time from $100\Gamma_0^{-1}$ to $10\Gamma_0^{-1}$ to $3\Gamma_0^{-1}$ ($\Gamma_{\mathrm{deph}}$=0). Encouragingly, for all of these values, and even for $\Gamma_{deph} = 0.3\Gamma_0$ or $\beta = 0.5$, the rise-time is significantly slower than the single-emitter response (c.f. white curve in Fig.~\ref{fig:RF_g2})), demonstrating that sub-radiance could be measured from the dynamics of the coupled system. In fact, as shown in Fig.~\ref{fig:RF_g2}c, it is only in the region where $\Gamma_{\mathrm{deph}} \approx 0.275 \Gamma_0$ and $\beta \approx 0.4$ that the rise-time approaches $\Gamma_0^{-1}$.

To further illuminate the role played by pure dephasing in the dynamics of the coupled system, we calculate the population of symmetric and asymmetric states, $\rho_\mathrm{ss}$ and $\rho_\mathrm{aa}$ (see Appendix~\ref{appen:sec1} for details), showing the results in Fig.~\ref{fig:rho}. Here, the emitters are again separated by $\Delta z_{12} = \pi/k$ and are initially in the ground state. At time $t = 0$ the pump is turned on and excite both emitters with the same local phase. As expected, in the case of no dephasing and at weak pumping $\Omega_{p,i} = 0.01 \Gamma_0$, we observe that the population of the sub-radiant state oscillates to a steady-state value $\rho_{\mathrm{sub}} = 0.5$ while the super-radiant state remains at $\rho_{\mathrm{sup}} = 0$.

A rapid loss in steady-state population in the presence of pure dephasing is seen in Fig.~\ref{fig:rho}b, where $\rho_{\mathrm{sub}}$ drops to 0.04 at $\Gamma_{\mathrm{deph}} = 0.1 \Gamma_0$. Interestingly, increasing the excitation intensity can help to mitigate the effects of dephasing.  That is, the imperfections of the coupled system can, to some degree, be mitigated through a careful selection of the excitation scheme.

A further interesting consequence of dephasing is observed in Fig.~\ref{fig:rho}b; even as $\rho_{\mathrm{sub}}$ decreases due to increased dephasing, $\rho_{\mathrm{sup}}$ increases. That is, dephasing, introduces a pathway between $\ket{a}$ and $\ket{s}$, allowing mixing between the two states. We observe that, for each excitation power,  $\rho_{\mathrm{sup}}$ peaks as dephasing increases and then gradually decreases. This demonstrates the balance between the increased mixing rate and loss of coherence introduced by the dephasing. The optimal excitation power for a given dephasing, as required to maximize $\rho_{\mathrm{sup}}$, is given in the inset. In the opposite scenario (not shown), where the system is initially pumped into the super-radiant state, we would then expect dephasing to provide a pathway to populate the sub-radiant state.

\subsection{\label{sec:phcwg}  Photonic crystal waveguides}
While many quantum photonic experiments employ the standard nanobeam waveguides described in the preceding sections, photonic-crystal waveguides (PhCWs) offer higher radiative coupling efficiencies \cite{Arcari:14}. The complex nanoscopic structure of PhCWs  modifies the flow of light and creates spectral regions where slow-light (i.e. with large group indices, $n_g$) modes enhance quantum light-matter interactions \cite{Lodahl:15,Hood:16}. The electromagnetic field is spatially modified with wavelength-scale feature sizes \cite{Javadi:18}. As a consequence, the Green's tensor is no longer of the form given by Eq.~(\ref{eq:Green1D}) and is obtained from numerical solutions to the field distribution emitted by a dipole $\bf{d}$ at position $\bf{r}'$ according to \cite{Novotny:06},
\begin{equation} \label{eq:G}
    \boldsymbol{\text{\textbf{E}}}\left(\bf{r}\right)=\omega^{2}\mu\mu_{0}  \bf{G}(\bf{r},\bf{r}',\omega_p)\cdot\bf{d}.
\end{equation}
Once the electric field distribution is known, the Green's tensor can be calculated using Eq.~\ref{eq:G}, instead of the simplified form given by Eq.~(\ref{eq:Ey}).

\begin{figure*}
    \includegraphics[trim={0 0cm 0 0},clip, width=0.9\textwidth]{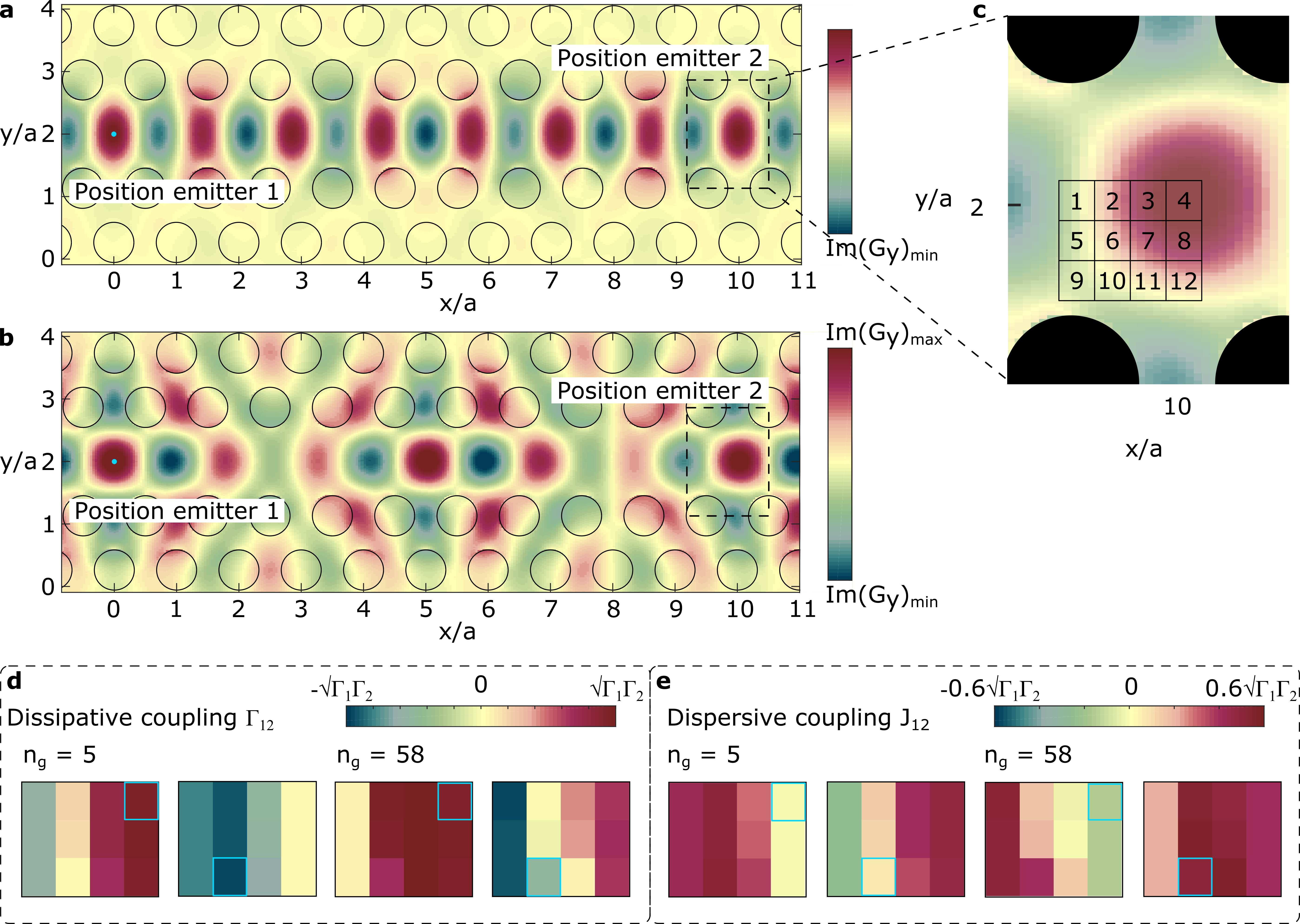}
  \caption{\label{fig:Gphcw} Components of the Green's tensor of a semiconductor  photonic-crystal waveguide. a) and b) show a spatial map of $\text{Re}\:\boldsymbol{G}_y\left(\boldsymbol{r},\boldsymbol{r}'\right)$, on the plane of symmetry in the middle of a PhCW membrane, calculated for a dipole emitter 1 at position $\boldsymbol{r}'$, for a fast- $\left(\text{n}_\text{g} = 5\right)$ and slow-light $\left(\text{n}_\text{g} = 58\right)$ mode. The axes are normalized to the lattice constant $a$, the air holes are shown as the black circles. The second emitter is located 10 unit cells away from emitter 1 in an area denoted by a dashed rectangle. c) Zoom in on the dashed area with possible emitter 2 positions numerated. Note that in a) and b) emitter 1 is placed at position 4 within its unit cell. d) and e) show the position dependent dissipative and dispersive coupling terms, $\Gamma_{12}$ and $J_{12}$, for two different group index values of $n_g = 5$ and $n_g = 58$. Each square displays the position dependent $\Gamma_{12}$/$J_{12}$ for emitter 2, where we consider two positions for emitter 1 which are outlined by a blue box.}
\end{figure*}

We compute the $\hat{\bf{y}}$ component of the Green's tensor numerically by calculating the field emitted by a $\hat{\bf{y}}$-oriented dipole embedded in a PhCW \cite{Javadi:18}, showing the results in Fig.~\ref{fig:Gphcw} for both a fast- $\left(\text{n}_\text{g} = 5\right)$ and slow-light $\left(\text{n}_\text{g} = 58\right)$ mode (see Appendix~\ref{appen:sec4} for more details).
As expected, the Green's tensor of the fast-light mode a) is confined to the center of the waveguide, while that of the slow-light mode b) is more delocalized. Note that these Green's tensor maps are for emitter 1 located at position 4 within the unit cell (c.f. Fig.~\ref{fig:Gphcw}c) at an anti-node of the mode. These were repeated at other positions for emitter 1, including position 10 away from the mode maximum and near an air hole.

Once the Green's tensor is known, $J_{12}$ and $\Gamma_{12}$ can be calculated according to Eqs.~(\ref{eq:J12}) and~(\ref{eq:Gamma12}), depending on the position chosen for emitter 2. Figure~\ref{fig:Gphcw} shows the results of such a calculation, for emitter 1 at positions 4 and 10 and emitter 2 at all positions, for both slow and fast-light modes, where $J_{12}$ and $\Gamma_{12}$ are normalised to the natural decay rates of emitters 1 and 2 ($\Gamma_1$ and $\Gamma_2$) within their respective unit cells, which varies depending on the position. These maps reveal a complex relationship between the dispersive and dissipative responses of the coupled system in the PhCW that depends on the position and group velocity in a way that is not present in the homogeneous nanobeam waveguides.

The coupling provided by the PhCW most closely resembles that of a nanobeam waveguide when emitter 1 is placed at the anti-node of the mode, seen as position 4 in Fig.~\ref{fig:Gphcw}c. For different positions of emitter 2 ($\mathrm{n}_\mathrm{g}=5$), we observe that $\Gamma_{12}$ and $J_{12}$ are out of phase, such that nearly perfect dissipative coupling can be achieved (left-most squares in Figs.~\ref{fig:Gphcw}d and e respectively). An exemplary case is when both emitters are placed at position 4, $\Gamma_{12} \approx \Gamma_1 (=\Gamma_2)$ and $J_{12} \approx 0$, meaning that near-perfect sub-radiance is possible, as we saw with the nanobeam waveguide (c.f. Figs.~\ref{fig:RF_g2} and~\ref{fig:rho}). Emitter 2 may be displaced, for example to positions 8 or 12, with only a slight reduction of the value of $\Gamma_{12}$, verifying  the robustness of PhCW. Alternatively, emitter 2 may be moved to positions 2, 6 or 10, in which case $\Gamma_{12} \approx 0$ and $J_{12} \approx 0.6\sqrt{\Gamma_1\Gamma_2}$, which tends towards a dispersively coupled system. A similar coupling behavior is observed when emitter 1 is placed away from the mode-maximum (i.e. position 10), albeit with reduced amplitudes and at different positions of emitter 2.

Interestingly, sub-radiant states may be easier to induce using slow-light modes ($\mathrm{n}_\mathrm{g}=58$). As we see in Fig.~\ref{fig:Gphcw}d and e (third square from the left), near-ideal dissipative coupling can still be achieved when emitter 1 is placed at position 4, while emitter 2 is in position 3 ($\Gamma_{12} \approx \sqrt{\Gamma_1\Gamma_2}$). Interestingly, displacing emitter 2 to positions 2 or 4 does not significantly alter $\Gamma_{12}$, yet it changes $J_{12}$ to nearly $\pm 0.2\sqrt{\Gamma_1\Gamma_2}$. Hence, by operating in the slow-light regime in a PhCW, one can lift the energy degeneracy of the super and sub-radiant modes. A similar but inverse trend is seen when emitter 1 is placed away from the mode maximum, near the air hole at position 10. Near-ideal sub-radiance is possible when emitter 2 is at positions 1, 5 or 9, yet in these scenarios we predict that $J_{12} \approx 0.2\Gamma_{0}$, again suggesting that a spectral separation between the super- and sub-radiant states is expected. These results indicate that PhCWs unlock the power of complex nanoscopic structures by realizing all effective emitter separations within one unit cell whilst maintaining a high coupling efficiency \cite{Arcari:14, Javadi:18}.

\section{\label{sec:conlusions} Conclusions}
In summary, we have theoretically studied the possibility of forming, exciting and detecting sub-radiant states using realistic (that is, imperfect) quantum emitters in nanophotonic waveguides. We find that dephasing rapidly deterioates signatures of sub-radiant states in spectral measurements, but that they persist in the dynamics observed in the photon statistics. We further showed that, even in the presence of dephasing, sub-radiant states may be efficiently populated if the emitters can be excited independently, although the dephasing leads to mixing between the super and sub-radiant states.

We have also numerically calculated the Green's tensor for a realistic photonic-crystal waveguide. We showed that with a careful selection of the position of the emitters, sub-radiance can be achieved while the energy degeneracy of the super- and sub-radiant states may be lifted. This study, combined with recent demonstrations of deterministic quantum-emitter photonic structure integration \cite{Sipahigil:16,Davanco:17,Schnauber:18,Pregnolato:20,Liu:21} provides a route to the controllable creation of sub-radiant states on-chip, providing a valuable resource to future quantum technologies.

\begin{acknowledgments}
The authors thank Anders Sondberg Sorensen, Sumanta Das, Matthew Foreman, and Daryl M Beggs for insightful discussions and Alisa Javadi for supplying the comsol simulation. The authors acknowledge financial support from Danmarks Grundforskningsfond (DNRF 139, Hy-Q Center for Hybrid Quantum Networks) and the EU's Horizon 2020 research and innovation programme (grant No. 824140, TOCHA, H2020-FETPROACT-01-2018). NR acknowledges funding from the Canadian Foundation for Innovation (CFI) and the Natural Sciences and Engineering Research Council of Canada (NSERC). 
\end{acknowledgments}

\appendix

\section{Hamiltonian and Lindblad operators}\label{appen:sec1}
To obtain an expression for the field operator, we evaluate the rising and lowering operators, $\hat{\sigma}^i_{eg}$ and $\hat{\sigma}^i_{ge}$ by turning to the master equation that describes the evolution of the atomic density matrix  $\hat{\rho}$ \cite{Martincano:11,Gonzaleztudela:11,Asenjogarcia:17}:
\begin{align}
\frac{d}{dt}\hat{\rho} = -\frac{i}{\hbar}[\mathcal{H},\hat{\rho}]+\mathcal{L},
\label{Eq:Master}
\end{align}
where $\mathcal{H}$ and $\mathcal{L}$ are the Hamiltonian and Lindblad operator respectively. In this paper, the results were calculated using the atomic density operator for two emitters written in the following basis:
\begin{eqnarray}
\hat{\rho} =\left(
\begin{array}{cccc}
\rho_{11} & \rho_{12} & \rho_{13} & \rho_{14} \\
\rho_{21} & \rho_{22} & \rho_{23} & \rho_{24} \\
\rho_{31} & \rho_{32} & \rho_{33} & \rho_{34} \\
\rho_{41} & \rho_{42} & \rho_{43} & \rho_{44} \\
\end{array}\right)\quad\text{and}\quad
\begin{array}{cc}
\ket{1}&=\ket{g_1g_2} \\
\ket{2}&=\ket{e_1e_2} \\
\ket{3}&=\ket{g_1e_2} \\
\ket{4}&=\ket{e_1g_2}, \\
\end{array}
\end{eqnarray}
where the subscripts 1 and 2 refer to emitter 1 and 2 respectively. Using this basis and some algebra, we can  write out the relevant atomic operators for emitter 1 and 2 in matrix form:
\begin{eqnarray}
\hat{\sigma}^1_{ge}= |g_1\rangle\langle e_1|=\left(\begin{array}{cccc}
0&0&0&1\\
0&0&0&0\\
0&1&0&0\\
0&0&0&0\\
\end{array}\right)\nonumber \\
\hat{\sigma}^1_{eg}= |e_1\rangle\langle g_1|=\left(\begin{array}{cccc}
0&0&0&0\\
0&0&1&0\\
0&0&0&0\\
1&0&0&0\\
\end{array}\right)\nonumber \\
\hat{\sigma}^2_{ge}= |g_2\rangle\langle e_2|=\left(\begin{array}{cccc}
0&0&1&0\\
0&0&0&0\\
0&0&0&0\\
0&1&0&0\\
\end{array}\right)\nonumber \\
\hat{\sigma}^2_{eg}= |e_2\rangle\langle g_2|=\left(\begin{array}{cccc}
0&0&0&0\\
0&0&0&1\\
1&0&0&0\\
0&0&0&0\\
\end{array}\right)\nonumber \\
\hat{\sigma}^1_{ee}= |e_1\rangle\langle e_1|=\left(\begin{array}{cccc}
0&0&0&0\\
0&1&0&0\\
0&0&0&0\\
0&0&0&1\\
\end{array}\right)\nonumber \\
\hat{\sigma}^2_{ee}= |e_2\rangle\langle e_2|=\left(\begin{array}{cccc}
0&0&0&0\\
0&1&0&0\\
0&0&1&0\\
0&0&0&0\\
\end{array}\right).\nonumber \\
\end{eqnarray}
Here we have assumed that each emitter is a two-level system with a ground state $|g\rangle$ and excited state $|e\rangle$, and the atomic operator describes the transition between these two states. 

We can obtain an effective Hamiltonian that describes our system within the Born-Markov approximation, which is valid when the relevant Green's function does not change over the emitter linewidth \cite{Hood:17}. This holds true for waveguides, except in photonic crystal waveguides close to the band-edge of the propagating mode \cite{Hood:16}. For a system of two dipoles in a waveguide, the effective Hamiltonian can be written as \cite{Asenjogarcia:17,Dzsotjan:10,Hood:17}:
\begin{eqnarray}
\mathcal{H} &= -\hbar\sum_{i=1,2}\Delta_i\hat{\sigma}^i_{ee} -\hbar\sum_{i,j=1,2} J_{ij}\hat{\sigma}^i_{eg}\hat{\sigma}^j_{ge} \nonumber\\
&-\hbar\sum_{i=1,2}(\hat{\Omega}_{p,i}^*\hat{\sigma}^i_{ge}+\hat{\Omega}_{p,i}\hat{\sigma}^i_{eg}),
\label{Hamiltonian}
\end{eqnarray}
where $\Delta_i = \omega_p-\omega_i$ is the frequency detuning between emitter $i$ and the probe field and $\hat{\Omega}_{p,i}= \boldsymbol{d}_i^*\cdot\hat{\boldsymbol{E}}_p^+(\boldsymbol{r_i})/\hbar$ is the guided mode Rabi frequency. The Hamiltonian also includes the dispersive coupling term $J_{12}$, which describes the energy level shifts due to the coherent waveguide-mediated dipole-dipole coupling and can be evaluated from the Green's function\cite{Asenjogarcia:17,Hood:17,Turschmann:17,Gonzaleztudela:11}, as detailed in the main text. We use the following Lindblad operator which contain the decay paths of the system:
\begin{eqnarray}
\mathcal{L}&=\sum_{i,j=1,2}\frac{\Gamma_{ij}}{2} (2\hat{\sigma}^i_{ge}\hat{\rho}\hat{\sigma}^j_{eg} -\hat{\sigma}^i_{eg}\hat{\sigma}^j_{ge}\hat{\rho} -\hat{\rho}\hat{\sigma}^i_{eg}\hat{\sigma}^j_{ge}) \nonumber\\
&+ \sum_{i=1,2}\frac{\Gamma_{i,gg}}{2} (2\hat{\sigma}^i_{gg}\hat{\rho}\hat{\sigma}^i_{gg} -\hat{\sigma}^i_{gg}\hat{\sigma}^i_{gg}\hat{\rho} -\hat{\rho}\hat{\sigma}^i_{gg}\hat{\sigma}^i_{gg})\nonumber\\
&+ \sum_{i=1,2}\frac{\Gamma_{i,ee}}{2} (2\hat{\sigma}^i_{ee}\hat{\rho}\hat{\sigma}^i_{ee} -\hat{\sigma}^i_{ee}\hat{\sigma}^i_{ee}\hat{\rho} -\hat{\rho}\hat{\sigma}^i_{ee}\hat{\sigma}^i_{ee}).
\label{Lindblad}
\end{eqnarray}
Here $\Gamma_{ii}$ is the natural decay rate of each emitter and $\Gamma_{i,gg}$, $\Gamma_{i,ee}$ are the pure dephasing rates of the ground and excited states respectively. In realistic solid-state systems, coupling to a phonon bath gives rise to pure dephasing processes which decrease the coherence of the emitter \cite{Lodahl:15}.

By evaluating the master equation, we obtain a set of rate equations that describe the evolution of the density matrix. The equations of motion of the density operator can be solved for the steady-state case, where we assume the system is excited by a continuous wave laser. However, in order to find the steady state solution, i.e. $\dot{\hat{\rho}}=0$, we first re-write the Hamiltonian and Lindblad operator in the so-called Dicke state basis \cite{Dicke:54}:
\begin{align}
\mathcal{H'} &= \mathbf{B}^{-1}\mathcal{H}\mathbf{B}\nonumber\\
L' &= \mathbf{B}^{-1}L\mathbf{B}\nonumber\\
\hat{\rho}'&=\mathbf{B}^{-1}\hat{\rho}\mathbf{B}\nonumber\\
\rightarrow\dot{\hat{\rho}}' &= -\frac{i}{\hbar}[\mathcal{H'},\hat{\rho}']+\mathcal{L'}
\end{align}
\begin{figure}
   \includegraphics[trim={0 0cm 0 0},clip, width=0.48\textwidth]{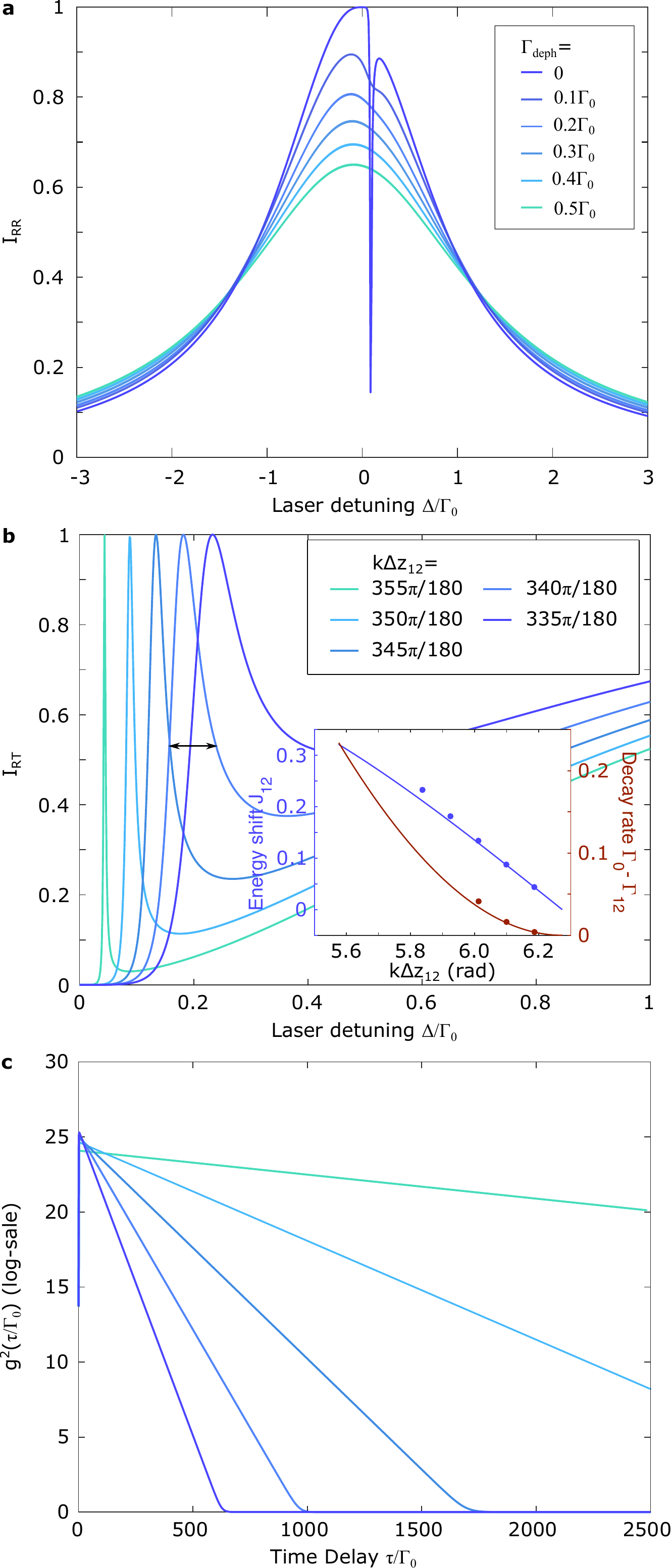}
   \caption{\label{fig:RT_deph_app} Sub-radiant features in intensity measurements in the presence of dephasing. a) Reflection spectra, $I_\mathrm{RR}$, for different $\Gamma_{\mathrm{deph}}$ for $k\Delta z_{12} = 35 \pi / 18$. The sharp sub-radiant feature, whose magnitude we label $\Delta T_\mathrm{sub}$, rapidly vanishes as $\Gamma_{\mathrm{deph}}$ increases, also shown in $I_{RT}$ in b) for different separations. Inset: the peak position follows the expected energy shift, $J_{12}$, and its width corresponds to the modified decay rate $\Gamma_0-\Gamma_{12}$. c) The corresponding RT $g^{(2)}(\tau)$ for the same separations $\Delta z_{12}$ show a large bunching peak at zero time delay. }
\end{figure}
The new basis now consists of symmetric and asymmetric intermediate states, which are entangled:
\begin{align}
\ket{g}&=\ket{g_1g_2}\nonumber \\
\ket{e}&=\ket{e_1e_2}\nonumber \\
\ket{s}&=\frac{1}{\sqrt{2}}(\ket{e_1g_2}+\ket{g_1e_2})\nonumber \\
\ket{a}&=\frac{1}{\sqrt{2}}(\ket{e_1g_2}-\ket{g_1e_2})
\end{align}
The steady state solution is hence obtained by setting $\frac{d}{dt}\hat{\rho}' = 0$ and solving the 16 coupled equations numerically. By numerically solving Eq. \ref{Eq:Master} to obtain the density matrix $\rho$, the time evolution of the populations of the entangled states $\ket{s}$ and $\ket{a}$, as shown in Fig.~\ref{fig:rho}, can also be acquired.

\section{Sub-radiance: dependence on dephasing and separation}\label{appen:sec2}
The reflection spectra, $\text{I}_\mathrm{RR}$, can be calculated using the same equation as $\text{I}_\mathrm{RT}$ and is displayed in Fig.~\ref{fig:RT_deph_app}a. The sharp sub-radiant feature vanishes even for moderate values of $\Gamma_\mathrm{deph}$. The emitter separation, $\Delta z_\mathrm{12}$, determines the ratio of dispersive coupling $J_{12}$ and dissipative coupling $\Gamma_{12}$, as is displayed in Fig.~\ref{fig:RT_deph_app}b. The sharp sub-radiant feature is primarily visible close to $k\Delta z = 2 \pi$, where the dissipative coupling dominates and hence giving rise to maximum sub-radiance, as seen in the inset. Similarly, a logarithmic decay at longer time scales is primarily a feature of the sub-radiance, which is also dependent on the different emitter separations.

\section{Rise-time fit} \label{appen:sec3}

For Fig.~\ref{fig:RF_g2}, we fit the individual $g^{(2)}(\tau)$ of the coupled emitter system to obtain the rise time. In order to compare it to the single-emitter antibunching, we use the following function \cite{Loudon:10}:
\begin{equation}
    f(\tau)=1-\frac{\Gamma e^{-\tau\Gamma/2}-\frac{\Gamma}{2}e^{-\tau\Gamma}}{\Gamma/2},
\end{equation}
where the rise time is given by the fit parameter $\Gamma$.

\section{Photonic crystal waveguide comsol simulation}\label{appen:sec4}

We performed a full 3D numerical simulation in Comsol Multiphysics based on the work in reference \cite{Javadi:18} to calculate the field emitted by a dipole over all space. This allows us to use Eq.~\ref{eq:G} to calculate the corresponding Green's tensor component for the PhCW. 

The PhCW details that we simulate is as follows: The refractive index was chosen to be 3.5, corresponding to the value for GaAs. The full simulation volume spans 34 unit cells in the x-direction and 17 rows of air holes (with refractive index 1) in the y-direction, where the unit cell and hole sizes are 240 nm and 160 nm respectively. The simulation method is briefly described as follows: First an eigenvalue calculation is performed to determine the eigenfrequency, eigenvector of the primary guided mode and the group index $\mathrm{n}_\mathrm{g}$, from which the correct boundaries are obtained.  A finite element frequency domain simulation is then carried out with a dipole in the PhCW using the correct boundary conditions. The simulation is repeated for $\mathrm{n}_\mathrm{g} = 5$ and $58$ and for different dipole positions within the unit cell. We consider the dispersion relation:
\begin{equation}
\mathrm{n}_\mathrm{g} = c/\mathrm{v}_\mathrm{g}=c/(\delta\omega/\delta k),
\end{equation}
where $\omega$ is the normalized frequency, k is the wave vector, $c$ is the speed of light in vacuum and $\mathrm{v}_\mathrm{g}$ is the group velocity that can be obtained by the slope of the guide mode. Finally, the dispersive and dissipative coupling terms are normalised to the individual decay rates of emitters 1 and 2 ($\Gamma_1$ and $\Gamma_2$) positioned in their respective unit cell.

\nocite{*}

%

\end{document}